\documentclass[11pt]{article}
\usepackage{bloisx,graphicx,amsfonts,hyperref,textcomp,eurosym}

\def\be{\begin{equation}}
\def\ee{\end{equation}}
\def\bea{\begin{eqnarray}}
\def\eea{\end{eqnarray}}

\usepackage{color}
\definecolor{IITred}{rgb}{0.5,0.05,0.05}
\definecolor{TrueBlue}{rgb}{0.05,0.05,0.5}
\definecolor{DrabGreen}{rgb}{0.05,0.5,0.05}

\newcommand{\gev}{\hbox{ GeV}}

\newcommand{\tev}{\hbox{ TeV}}

\newcommand{\nb}{\hbox{ nb}}
\newcommand{\pb}{\hbox{ pb}}
\newcommand{\fb}{\hbox{ fb}}

\newcommand{\lum}{\hbox{ cm}^{-2}\hbox{ s}^{-1}}

\DeclareMathSymbol{\lll}          {\mathrel}{AMSa}{"6E}

\def\ltap{\mathop{\raisebox{-.4ex}{\rlap{$\sim$}} 
\raisebox{.4ex}{$<$}}}
\def\gtap{\mathop{\raisebox{-.4ex}{\rlap{$\sim$}} 
\raisebox{.4ex}{$>$}}}

\def\abs#1{\left| #1\right|}

\def\slashii#1{\setbox0=\hbox{$#1$}             % set a box for #1
   \dimen0=\wd0                                 % and get its size
   \setbox1=\hbox{\sl/} \dimen1=\wd1            % get size of /
   \ifdim\dimen0>\dimen1                        % #1 is bigger
      \rlap{\hbox to \dimen0{\hfil\sl/\hfil}}   % so center / in box
      #1                                        % and print #1
   \else                                        % / is bigger
      \rlap{\hbox to \dimen1{\hfil$#1$\hfil}}   % so center #1
      \hbox{\sl/}                               % and print /
   \fi}                                         %
\def\slashiii#1{\setbox0=\hbox{$#1$}#1\hskip-\wd0\hbox to\wd0{\hss\sl/\/\hss}}
\newcommand{\tp}{\textit{These Proceedings}}

\begin{document}
\vspace*{4cm}
\title{HIGHLIGHTS \& PERSPECTIVES: \\ XXII Rencontres de Blois}

\author{CHRIS QUIGG}

\address{Theoretical Physics Department, Fermi National Accelerator Laboratory\\
P.O. Box 500, Batavia, Illinois 60510 USA}

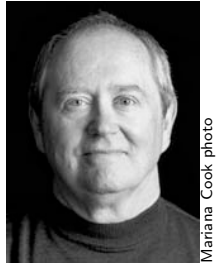
\maketitle\abstracts{
 This is a brief digest of my \href{http://confs.obspm.fr/Blois2010/Quigg.pdf}{closing lecture} at the XXII Rencontres de Blois, \textit{Particle Physics and Cosmology.} Slides of all the talks referred to may be found at \url{http://confs.obspm.fr/Blois2010}. \hfill \textsf{FERMILAB-CONF-10-367-T}
 }

\section{Historique \label{sec:hist}}
It is a tradition of the \textit{Rencontres de Blois} to blend scientific discussions at the highest level with culture in an atmosphere that makes us aware of our heritage, broadly conceived. This splendid chateau and the art and history revealed during the excursion to Chenonceau and Clos Luc\'{e}  have enhanced our experience this week. But, in this region so rich in patrimony, there is more: two of the most celebrated historical figures \textit{d'origine Bl\'{e}soise} had connections with physics, and so may be counted among our scientific ancestors!

If you have ventured to the far end of the esplanade, you will have noticed the \textit{Mus\'{e}e de la Magie}~\cite{museemagie}, a monument to the life and work
of Jean-Eug\`{e}ne Robert-Houdin (1805--1871), the most famous magician in all of France and
the father of modern conjuring. Robert-Houdin was the
first magician to appear in formal wear, and the first to use electricity in his act. Part of his
legacy is a posthumous work~\cite{roberth} devoted to magic and ``amusing physics.''

The second famous son of Blois is Denis Papin (1647-1714), the master of steam
power who worked in London with Robert Boyle. In 1679, he invented the  \textit{marmite de Papin}~\cite{denisp}, \textit{cocotte minute,} or  pressure cooker,  a kind of cooking-pot in which arbitrarily tough meat can be rendered soft. Papin thereby established what we recognize today as the Standard Model of English Cuisine.  

\section{Arrival of the LHC \label{sec:lhc}}
The signal event of this year in particle physics is the arrival of the Large Hadron Collider at CERN as a research instrument. We celebrate both the performance of the collider itself, described by Lucio Rossi~\cite{rossi}, and the impressive early analyses carried out by the experimental teams. One measure of the machine development is that, at the time we met in Blois in July, the LHC had delivered more than $350\nb^{-1}$ to the ATLAS and CMS detectors, at $3.5\tev$ per beam. By September 6, the integrated luminosity was an order of magnitude higher~\cite{lhclum}, and the peak luminosity had surpassed $10^{31}\lum$. The goal for the 2010-2011 run is to accumulate $\approx 1\fb^{-1}$ at $\sqrt{s} = 7\tev$. The current expectation for post-shutdown operation is to reach $\sqrt{s} = 13\hbox{ - }14\tev$ during 2013, and to progress toward the design luminosity of $10^{34}\lum$. CERN's conception of the future, described by Rolf Heuer~\cite{rolf}, includes options for a high-luminosity LHC, an electron-positron collider, and more. The European strategy for particle physics should be revisited in 2012.

Members of the experimental collaborations have reported here on their experience in commissioning and calibrating the detectors~\cite{LHCcom}, and on some early analyses~\cite{LHCearly}. To note just a few of the interesting results, I cite ALICE's studies of particle production, including measurements of the charged multiplicity~\cite{yves}, CMS measurements of two-particle correlations~\cite{roeck}, ATLAS investigations of high-transverse-momentum jets~\cite{margie}, and the observation of sequential decays of heavy quarks by LHC$b$~\cite{andreas}.  It is exciting to hear that LHC$b$ will begin to confront D0's surprising dimuon charge asymmetry~\cite{Abazov:2010hv} at an integrated luminosity of $\approx 100\pb^{-1}$. Many more results were available one week after Blois, at ICHEP 2010 in Paris~\cite{ichepparis}. All of this testifies to the skillful planning and execution by accelerator physicists and experimenters, and also to their dedication and stamina!

For all of this impressive progress, we cannot ignore the fact that the LHC is, for now, operating at only half its design energy. This has implications for the rate at which we might anticipate discoveries, and also for relative advantages of the experimental campaigns at the LHC and Tevatron. Parton luminosities, supplemented by what we know from measurements at the Tevatron and from simulations for the 14-TeV LHC, are a reliable tool for assessing what we can expect~\cite{Quigg:2009gg}. Although every measurement or search is a special case for which we must consider both signals and backgrounds, as a general rule the LHC experiments begin to open terrain not explored by the Tevatron when their integrated luminosity exceeds a few hundred$\pb^{-1}$. Guido Altarelli reviewed the TeV-scale physics setting~\cite{guido} and the prospects for new insights early in the life of the LHC. Among many possibilities, I regard the discovery of a diquark resonance~\cite{Bauer:2009cc} (for which the $pp$ collisions of the LHC offer higher sensitivity than the $\bar{p}p$ collisions of the Tevatron) as not so plausible, but the early observation of a fourth-generation quark~\cite{4thgen,soni} as not so implausible. 

\section{The Tevatron in Its Prime\label{sec:teva}}
Meanwhile, across the Atlantic, the Tevatron has never been in better form: the peak luminosity has reached $4 \times 10^{32}\lum$ and more than $9\fb^{-1}$ has been delivered to CDF and D0~\cite{tevlum}, which are reporting new results on many topics. To indicate the breadth of scientific interests, I note a D0 study of $\bar{p}p$ elastic scattering~\cite{d0el}, which manifests the expected shrinkage of the diffraction peak. This is just one among many Tevatron results on the strong interactions~\cite{tevqcd}. More generally on strong interactions, we heard a comprehensive account of QCD by Varelas~\cite{varelas}; an update on the HERA parton distribution functions~\cite{glazov}; and progress reports on investigations of the quark-gluon-plasma~\cite{qgp} and of hot and dense baryonic matter. Holographic techniques informed by gauge-gravity duality open the investigation in the strong-coupling regime of theories that are not QCD, but might have properties in common with QCD~\cite{hashi}.

Among searches, we may mention the $t^\prime$ search from D0~\cite{d0tpr} and the $Z^\prime$ search from CDF~\cite{cdfzp}, which limits  $M(Z^\prime_{\mathrm{SM}}) > 1\,071\gev$ using $4.6\fb^{-1}$. For the evolution of $Z^\prime$ searches from the Tevatron to the LHC, see~\cite{zptevlhc}, and for other new physics searches, see~\cite{tevnp}.

The Tevatron experiments have established an enviable record for precise measurements. The combined Tevatron top-quark mass is now known so precisely, $m_t = (173.3 \pm 0.6\hbox{ (stat)} \pm 0.9\hbox{ (syst)})\gev = (173.3 \pm 1.1) \gev$~\cite{shaba}, that it is urgent to confront the question of  exactly what quantity is measured~\cite{Smith:1996xz}. We have also heard about new determinations of top-quark properties~\cite{topprop} and electroweak observables~\cite{ewprog}.

Encouraged by the increasing incisiveness of their Higgs-boson searches~\cite{daniela,tevhpar}, the Tevatron collaborations have proposed to extend running beyond the planned October 2011 cutoff, and to continue for three more years, to accumulate $\sim 16\fb^{-1}$ for analysis. At ICHEP, CDF and D0 excluded a standard-model Higgs boson in the range $158\gev \ltap M_H \ltap 175\gev$ (and also $100\gev \ltap M_H \ltap 109\gev$) at 95\% CL~\cite{ichepH}. The reduced energy of the current LHC run and the LHC shutdown for retrofitting in 2012-2013 also enter into a reassessment of the Tevatron's potential to contribute to the investigation of electroweak symmetry breaking. The centerpiece of the proposal is that at $16\fb^{-1}$ the combined-experiments / combined-channels sensitivity for the standard-model Higgs boson would exceed 3-$\sigma$ ``evidence'' for $100\gev \ltap M_H \ltap 185\gev$~\cite{fnalpac}. Although a decision to continue running would be nontrivial because of budgetary constraints and the interaction with Fermilab's future program, the Physics Advisory Committee has responded with considerable enthusiasm~\cite{fnalpacaug}. The verdict rests with the laboratory management and the funding agencies.

\section{Learning to See at the LHC \label{sec:L2C}}

The LHC is beginning to advance the experimental frontier of particle physics to the heart of the TeV scale, where we are confident that we will find new insights into the nature of electroweak symmetry breaking. \textit{We do not know what the new wave of exploration will find.} Precisely because we do not know the answers, it is imperative to look broadly, and this has been the thrust of many plenary talks~\cite{theorists} and contributions in the parallel sessions~\cite{LHCpar}. Along with our conviction that exploration of the 1-TeV scale will give definitive answers about electroweak symmetry breaking, we have good reason to hope that we might also find candidates for the dark matter of the universe and resolve the hierarchy problem.

I believe that we should also take advantage of the opportunity to learn more about the richness of the strong interactions, especially in the realm of ``soft'' particle production that theorists cannot describe by controlled calculations in perturbation theory. I would like to emphasize that the object of initial studies is not merely to tune PYTHIA parameters (which may not have direct physical significance)~\cite{pytune}, as useful as that exercise may be. The first conclusion of the LHC experiments is that the pre-LHC event generators did not perfectly anticipate what was observed.

Experimental studies in the 1970s established the essential features of multiple production at energies up to $\sqrt{s} = 63\gev$: Feynman scaling, with distinct diffractive and ``multiperipheral'' components, the latter characterized by short-range order in rapidity. 
This doesn't mean that we can regard ``soft'' particle production as settled knowledge. Tevatron studies have been informative but not exhaustive, so we can't be sure that what was learned in the 1970s accounts for all the important features at Tevatron energies and beyond. At the highest energies, well into the ($\propto \ln^2{s}$?) growth of the $pp$ total cross section, long-range correlations might show themselves in new ways. {The high density of partons carrying  $p_z = 5\hbox{ to }10\gev$ may give rise to hot spots in the spacetime evolution of the collision aftermath, and thus to thermalization or other phenomena not easy to anticipate from the QCD Lagrangian.} We might anticipate a growing rate of multiple-parton interactions, perhaps involving correlations among partons: the quark-diquark component of the proton might manifest itself in elementary collisions involving diquarks. The $\ln{s}$ expansion of the rapidity plateau softens kinematical constraints in the central region, and the sensitivity to high-multiplicity events  (or otherwise rare occurrences) of modern experiments vastly exceeds what could be seen with bubble-chamber statistics. 
For all these reasons, I suspect that a few percent of minimum-bias events collected at $\sqrt{s} \gtap 2\tev$ might display unusual event structures. 

\textit{Looking at events} can play an important role~\cite{Quigg:2010nn}, not only to refine our intuition, but also to discover candidate new physics that might become the object of dedicated study in the future. Because I expect the event structure to evolve with increasing collision energy, I would like to see during 2010-2011 \textit{a set of modest dedicated runs at steps in energy, e.g., at} $\sqrt{s} = 0.9, 2, 3.5, 5, 7\tev$, lightly triggered, to survey the nature of particle production. Now that the essential performance of the detectors has been validated, such a survey would be well worth the disruption it would cause to routine operations and the accumulation of integrated luminosity at $7\tev$. We should use this first LHC run to learn what we will want to study in depth beginning in 2013.

\section{Neutrinos}
The investigation of neutrino properties and interactions, and the search for extraterrestrial sources of neutrinos, was also well-represented in Blois~\cite{nuprog}. Among new initiatives, which include the start-up of the T2K program and fresh data from ANTARES, we welcome the first $\nu_{\mu} \to \nu_{\tau}$ candidate from OPERA~\cite{Agafonova:2010dc}. This specimen puts a face on the inference that $\nu_{\mu} \leftrightarrow \nu_{\tau}$ mixing is the dominant phenomenon in atmospheric neutrino oscillations~\cite{Ashie:2005ik}. Review talks by Kayser~\cite{boris} and Wojcicki~\cite{stanw} summarized the current status and recent experimental progress.

The MINOS collaboration has reported disappearance results from their first antineutrino run~\cite{minos}; the antineutrino mixing angle and mass-squared difference are in some tension with the corresponding neutrino values. While this may well be a transitory effect of modest statistics, it is worth stretching our minds on possible implications---especially those less radical than CPT violation. For example, it is worth asking whether nonstandard interactions that survive other experimental sieves could give rise to an apparent difference in $(\sin^22\theta,\Delta m^2)$~\cite{belen,Kopp:2010qt}.

A decade of progress in neutrino mixing leaves us with a great many other unanswered questions, including: Do neutrino masses display a normal or inverted spectrum? What is the value of the small mixing angle, $\theta_{13}$?
Are neutrinos their own antiparticles?
How many mass eigenstates are there?
Are sterile neutrinos required to understand neutrino mixing?
Do neutrinos have any peculiar lectromagnetic properties?
Is CP symmetry respected in neutrino interactions?
Do some or all of the light neutrinos experience nonstandard neutrino interactions?
What is the origin of neutrino mass?
What new surprises might nature have in store for us? All that we need to know has stimulated many promising new initiatives in neutrino physics~\cite{y2k}.

\section{Quark Flavor Physics \label{sec:qflav}}
The rich field of quark flavor physics was also strongly represented in the results presented at Blois~\cite{qflav}, along with new determinations of tau-lepton properties~\cite{tau}, and studies of hadron physics~\cite{hadron}. Christian Kiesling summarized the state of our knowledge on CP violation~\cite{kiesling}. Having established the outlines of flavor physics---charged-current interactions that exhibit, to good approximation, a three-generation $V-A$ form; the suppression of flavor-changing neutral currents by the GIM mechanism; and CP-violating phenomena prescribed by the CKM quark-mixing matrix---we now must take a deeper look, to see just how accurately our standard-model idealization conforms to reality.

On closer examination, experiments indicate a number of anomalies at a provocative level of statistical significance~\cite{soni}. Among these, the inclusive-exclusive tension in determinations of the quark-mixing-matrix element $\abs{V_{ub}}$ is notable for its persistence ~\cite{roney}. It is amusing (at least) that the discordant determinations of $\abs{V_{ub}}$ from inclusive decays ($B \to X_u \ell \nu$), the exclusive decays $B \to \pi \ell \nu$ (mediated by the vector current), and the annihilation decay $B \to \tau \nu_\tau$ (mediated by the axial current) could be reproduced if a small right-handed charged-current interaction were present~\cite{Buras:2010pz}.

Determinations of the CP-violating phase $\beta_s^{J\!/\!\psi\phi}$, measured in $B_s \to J\!/\!\psi\phi$ decays, have previously differed (at $2.1\sigma$) from standard-model expectations~\cite{cdfd0bph}. The disagreement is somewhat mitigated by a new CDF measurement that benefits from increased statistics and improved $B$ tagging~\cite{cdfbphase}. Forthcoming data, together with measurements of the lifetime difference between the two $(B_s,\overline{B}_s)$ mass eigenstates, will probe for new physics. Indeed, $B$ physics promises much new information from BaBar and Belle, CDF and D0, and the LHC experimentsl~\cite{lhcflav}. New initiatives are in the works as well~\cite{comingatt}; in particular, funding in the amount
of $10^8$\yen\ $\approx$ \EUR{87M} over the next three years
has been secured for the KEKB upgrade. We may look forward to a new round of $e^+e^-$ collisions beginning April 1, 2014!
 
\section{Cosmic Issues \label{sec:cosmic}}
We begin with two reports on ``conventional'' astrophysics.  We were treated to a remarkable example of modern observational capabilities in the reconstruction of stellar orbits around the massive black hole at the center of our galaxy~\cite{liebundgut}. Steady progress toward the detection of gravitational waves raises the fascinating prospect of a future in which the (non)observation of gravitational waves may serve as diagnostics for astrophysical phenomena~\cite{bauer}. 

Astro/Cosmo/Particle physics is rich in new results and the prospect of others, thanks to many new instruments and their planned successors~\cite{astrocont}. Over the past three decades, the hot big-bang cosmology established in the 1960s has been revised several times to incorporate new ideas and new observations. As provisional as it must be, the concordance cosmology, including an inflationary epoch and the introduction of dark matter and some form of dark energy, has a pleasing economy and consistency~\cite{malik}. A central question is whether the dark energy is a manifestation of a cosmological constant, or has a dynamical origin~\cite{khoury}. An audacious proposal to study dark energy over a range in redshifts by measuring $dz/dt$ is an element of the E-ELT Project~\cite{eelt}.

A decade after the discovery that the universe is expanding at an accelerating rate, it is worth restating how remarkable are the implications of the cosmological-constant interpretation. Not only are we living during an epoch in which the matter and dark-energy densities are comparable, we are at the threshold of a new inflationary age. It is worth bearing in mind that the inflationary $\Lambda$CDM cosmology is less a coherent theoretical framework than an assembly of modules---ideas and inventions---added in response to observations~\cite{Steinhardt:2004gk}. It is not yet grounded in general principles, so it is important that we remain skeptical, probe the idealizations, look for deviations, and seek a more holistic foundation.

A wealth of observational information from the cosmic microwave background, baryon acoustic oscillations, and the Union08 supernova data set points to a universe that is to excellent approximation flat, but dominated by something other than matter~\cite{Kowalski:2008ez,slosar}. Indeed, the latest W-MAP analysis, within the concordance cosmological model, points to a mass-energy budget of the present universe consisting of about $73\%$ dark energy, $23\%$ dark matter, and $4.5\%$ normal atomic matter~\cite{hinshaw}. It was not ever thus: according to the standard thermal history of the universe, at the surface of last scattering (age ca. $380\,000$ years), the universe was made up of roughly $63\%$ dark matter, $15\%$ photons, $12\%$ ordinary baryonic matter, and perhaps $10\%$ neutrinos. Dark energy was a trace component. The Planck satellite has completed its initial observing campaign in excellent form; we eagerly await the first wave of analyses, as well as the results of projects now in progress to measure polarizations of the cosmic microwave background radiation~\cite{paolod}.

Although we must always be alert to the possibility that we have misread the evidence, it is virtually certain that collisionless cold dark matter is a significant component of the present universe. Crafty observational strategies, such as the analysis of gravitational lensing of distant light sources that produced the COSMOS 3-dimensional map~\cite{cosmos}, begin to tell us where the dark matter resides. We know what it is not: neither baryonic nor, for the most part, neutrinos. We don't know how many species, and we should resist jumping to the conclusion that there is only one dark matter.

The quest for dark matter as weakly interacting massive particles has three main elements: the direct detection of thermal relics from the early universe by passive experiments ~\cite{dmcont}; indirect searches that aim to observe (co-)annihilation products of dark-matter particles with cosmological lifetimes~\cite{dmind}; and collider searches that could characterize the properties of dark-matter candidates in great detail, but of course cannot establish cosmological lifetimes. It remains a possibility that some or all of the dark matter is in the form of axions~\cite{axion}, which could guide us to a solution of the strong CP problem.

On the collider front, it is tantalizing that many proposals for physics beyond the standard model lead to dark-matter candidates, and that a relic density of the right magnitude naturally arises for WIMP masses that lie between $100\gev$ and $1\tev$, the range accessible to the Tevatron and LHC. With respect to the detection of thermal relics or their annihilation products, many new techniques and instruments are reporting results over an interesting range of masses and interaction cross sections~\cite{bernard}. For all the approaches, the experimenter must confront these questions: Is there a signal? If yes, is it background? If no, prove you are sensitive. The next five years promise a lot of excitement!

\section{The Quy Nhon International Center for Interdisciplinary Science Education}
On Monday afternoon the founding father of the \textit{Rencntres de Blois}, Tr\^{a}n Thanh V\^{a}n, presented his vision of a new international science and education center, to be built in Vietnam~\cite{van}. It is an ambitious plan---some might say, extravagant---and a natural response is, ``This time, he's gone too far.'' I venture to speculate that many of Van's past achievements, all the way back to the \textit{Rencontres de Moriond,} might have elicited that same response, when first exposed. In his \textit{Elegy} to Robert Lowell, the Irish master Seamus Heaney writes, ``The way we are living, timorous or bold, will have been our life.'' Let us follow our friend Van's example and, in science and in life, be bold!

\section*{Acknowledgments}
Fermilab is operated by the Fermi Research Alliance under contract number DE-AC02-07CH11359 with the U.S. Department of Energy. I acknowledge with pleasure the generous support of the Alexander von Humboldt Foundation, and thank Andrzej Buras for a warm welcome at Technische Universit\"{a}t M\"{u}nchen, where this report was completed. I owe special thanks to Liz Simmons and Boris Kayser for assisting in the preparation of my talk.

Pour terminer, je souhaite un tr\`{e}s grand merci  \`{a} tous les participants, aux gentils organisateurs des Rencontres de Blois, \`{a} nos amies sauvetrices du secretariat, au personnel du Ch\^{a}teau de Blois,  \`{a} Kim et Van.

\end{document}